\PassOptionsToPackage{numbers,sort&compress}{natbib}
\documentclass[3p,twocolumn]{elsarticle}
\journal{Energy}
\usepackage{lineno}
\usepackage{natbib}
\begin{document}
\begin{frontmatter}

\title{Improving electric power generation of a standalone
	wave energy converter via optimal electric load control}

\tnotetext[t1]{Accpeted by Energy}
\tnotetext[t2]{Email address:wanglg7@mail.sysu.edu.cn}

\author[add1,add2]{LiGuo~Wang}

\author[add1]{MaoFeng~Lin}
\author[add3,add4]{~Elisabetta~Tedeschi}
\author[add5]{~Jens~Engström}
\author[add5]{~Jan~Isberg}

\address[add1]{School of Marine Engineering and Technology, Sun Yat-Sen University, 519082 ZhuHai, China}
\address[add2]{Southern Marine Science and Engineering Guangdong Laboratory, 519082 ZhuHai, China}
\address[add3]{Department of Electric Power Engineering, Norwegian University of Science and Technology, N-7034 Trondheim, Norway}
\address[add4]{Department of Industrial Engineering, University of Trento, Trento, Italy}
\address[add5]{Department of Electrical Engineering, Uppsala University, 75121 Uppsala, Sweden.}

\begin{abstract}
This paper aims to investigate electric dynamics and improve electric power generation of an isolated wave energy converter that uses a linear permanent magnet generator as the power take-off system, excited by regular or irregular waves. This is of significant concern when considering actual operating conditions of an offshore wave energy converter, where the device will encounter different sea states and its electric load needs to be tuned on a sea-state-to-sea-state basis. To that end, a fully coupled fluid-mechanical-electric-magnetic-electronic mathematical model and an optimization routine are developed. This proposed time-domain wave-to-wire model is used to simulate the hydrodynamic and electric response of a wave energy converter connected to specific electric loads and also used in an optimization routine that searches optimal resistive load value for a wave energy converter under specific sea states. Sample results are presented for a point-absorber type wave energy converter, showing that the electric power generation of a device under irregular waves can be significantly improved. 
\end{abstract}



\begin{keyword}
wave energy converter, wave power system, wave-to-wire model, linear permanent magnet generator, electric load optimization.
\end{keyword}
\end{frontmatter}

\section{Introduction}
\label{sec.introduction}
Ocean wave energy is a promising renewable source to contribute to supply the world's energy demand, with an estimated world-wide potential of $2-3 TW$ \cite{falnes_review_2007}. Recent research on wave energy utilization started in the 1970s against the backdrop of the oil crisis, as reviewed in \cite{falcao_wave_2010}. In past decades, numerous coastal countries, such as the U.K., China, Sweden, Norway, the U.S.A, France, Italy, Ireland, Japan and Australia, have been making considerable efforts to develop wave energy conversion technologies. As outcomes, various concepts and approaches have been proposed to capture the energy carried by ocean waves, among them, the feasibility of some devices were examined by analytical, numerical and experimental methods, as reviewed in \cite{falnes_review_2007,ekstrom_electrical_2015,lopez_review_2013,clement_wave_2002,JIN2019450,LAVIDAS2020117131}. Results indicate that, the energy carried by ocean waves can be effectively captured and converted by wave energy converters (WECs) in a variety of ways. This energy can be used for different purposes, e.g., providing electricity, heating water, powering desalination plants or enabling transportation, with a minimal negative influence on the environment. However, to date, the majority of those devices are at a pre-commercial stage, and most WECs that aim to provide electricity are off-grid standalone devices.

A key performance indicator for an isolated full-scale WEC operated in open sea is the system efficiency, as it determines the power performance and dominates the economic performance of a device. Considerable work has been conducted to improve system efficiency, by employing control strategies \cite{korde_control_2000,wang_review_2018,hong_review_2014} for the power take-off (PTO) system or by optimizing geometric design of the primary capture system \cite{sjokvist_optimization_2014,kurniawan_optimal_2013,shadman_geometrical_2018,garcia-rosa_control-informed_2015,wen_shape_2018,yucheng14wec,timedomain,powerabsorption}. However, most endeavours so far have been limited to carrying out them independently, ignoring the coupling between subsystems or the losses, which are crucial for the accurate investigation of electric dynamics and power performance of a WEC and to the accurate determination of optimal linear electric load that maximizes the generated power. More important, most of the developed control strategies are intended to improve the power absorbed from ocean waves, rather than the electric power generated, which is not appropriate for the non-ideal cases, e.g, considering power transmission losses or non-ideal power take-off systems. The numerical solution to address that is to develop a mathematical model that incorporates all the components, from ocean wave side to electric grid or electric load side, whereas the physical solution is to perform tank or open-sea experiments. Considering the high capital and time cost of physical experiments, numerical simulation with a time-domain mathematical model, referred to as the $wave-to-wire$ model, is promising for studying system dynamics and energy losses, and identifying optimal electric loads. Additionally, wave-to-wire models can be used in model-based control of a WEC in regular and irregular waves, where the control strategies are used to improve device performance.

\begin{figure}[!htbp]
    \centering
    \includegraphics[width=7.8cm]{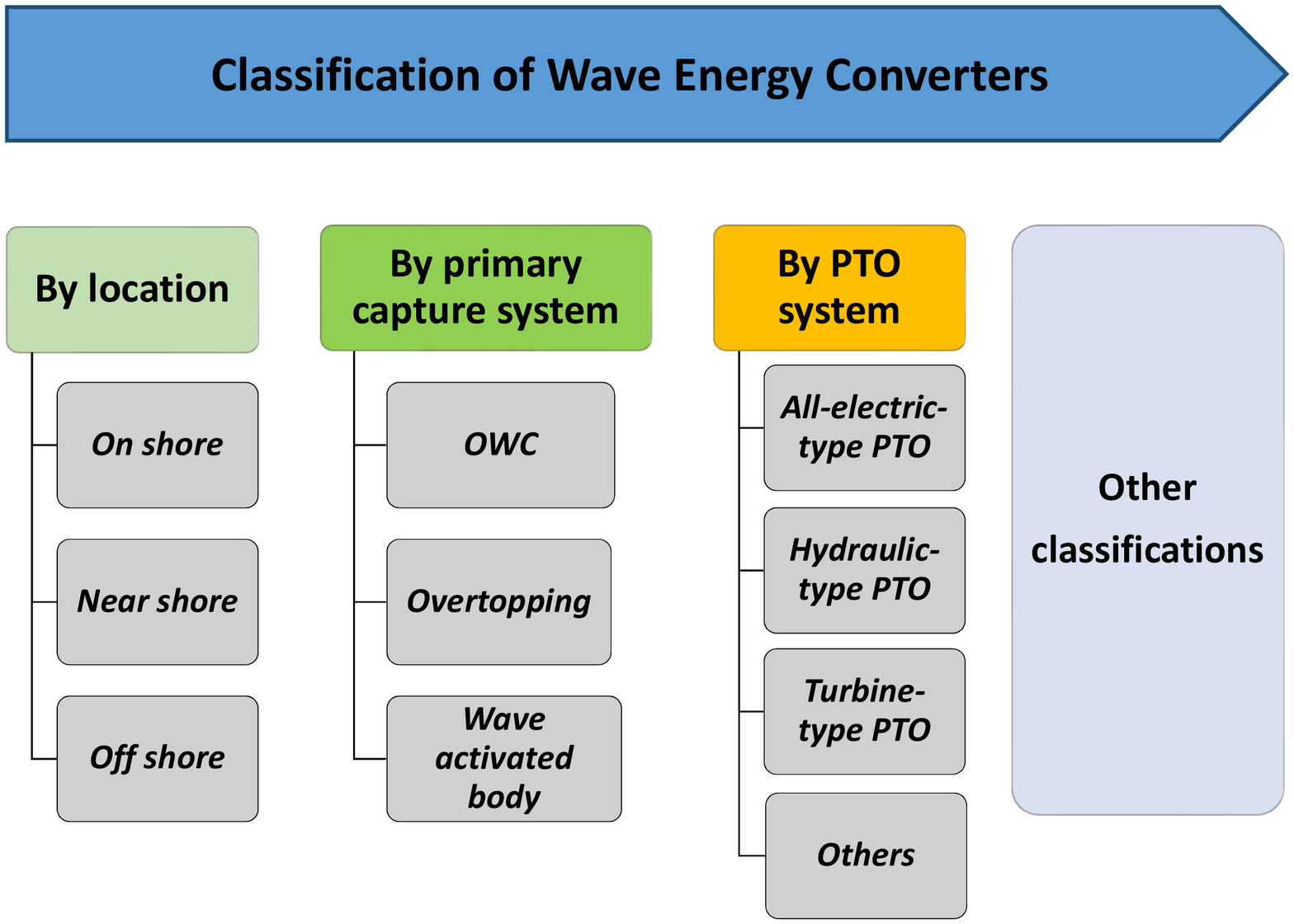}
    \caption{Classification of WECs by different criteria\cite{wang_review_2018}.}
    \label{fig:wecclassification}
\end{figure}

Several wave-to-wire models have been proposed for different types of WECs (a classification of WECs is presented in Fig. \ref{fig:wecclassification}). In \cite{PENALBA2019367}, two wave-to-wire models, including one with a variable-pressure hydraulic PTO and one with the constant-pressure configuration are presented. In \cite{babaritwave2wire}, a wave-to-wire model, where the radiation force memory term is replaced by additional states using the Prony method in order to formulate the equation of motion into the ordinary differential equation form, is presented for a WEC called SEAREV that employs a hydraulic PTO. In \cite{forehand_fully_2016}, a wave-to-wire time-domain model is used to simulate the mechanical, hydrodynamic, and electric response of an array of WECs, and the radiation force memory terms are replaced by the state-space methods. In \cite{BAILEY2016248}, a comprehensive numerical time-domain model is presented for an oscillating water column (OWC) WEC that uses a turbine, to provide accurate prediction of power generation, and a reduced-order model is presented in \cite{SUCHITHRA2019614}. In \cite{lifesaver2013}, a wave-to-wire model is proposed for a WEC called Lifesaver that employs a rotary permanent magnet generator, and the key result from simulations is that when taking generator losses into account, the performance of reactive control is not satisfactory. Recent reviews of the wave-to-wire work can be found in \cite{wang_review_2018,penalba_review_2016}, where WEC control issues are discussed and summarized. 

The permanent magnet WEC concept used in this paper is a point-absorber type wave energy converter (PAWEC) with a linear permanent magnet generator (LPMG), as explained in the earlier publications from the wave energy research group at Uppsala University\cite{bostrom_experimental_2010}. To the best of authors' knowledge, no completed work has been conducted to simulate the electric dynamics and the power conversion of this WEC with optimal electric load, in particular for irregular wave cases. Another novel work is that an optimization procedure is developed to find the optimal resistor load value for the device excited by irregular incident waves, based on the proposed wave-to-wire time-domain model. Additionally, power losses of the electric power cable are included in the modeling and optimization process.  

The structure of this paper is as follows. Section \ref{sec:methods} presents the optimization methodology used for finding the optimal electric load for a PAWEC that employs a linear permanent magnet generator as the PTO system. Section \ref{sec:model} explains the wave-to-wire model in detail, including the fluid-structure interaction theory, the linear generator model, the power electronic converter model, and electric load model. In Section \ref{sec:casestudy}, the developed wave-to-wire model and optimization routine is applied to a floating PAWEC, originally from a full-scale WEC developed by Uppsala University, and its electric performance is simulated and the optimal electric load is identified. Section \ref{sec:discussion} presents a discussion based on simulation results from case studies. Summative conclusions are drawn in Section \ref{sec:conclusion}.
\section{Methodology}
\label{sec:methods}
\begin{figure}[!ht]
	\centering
	\includegraphics[width=7.8 cm]{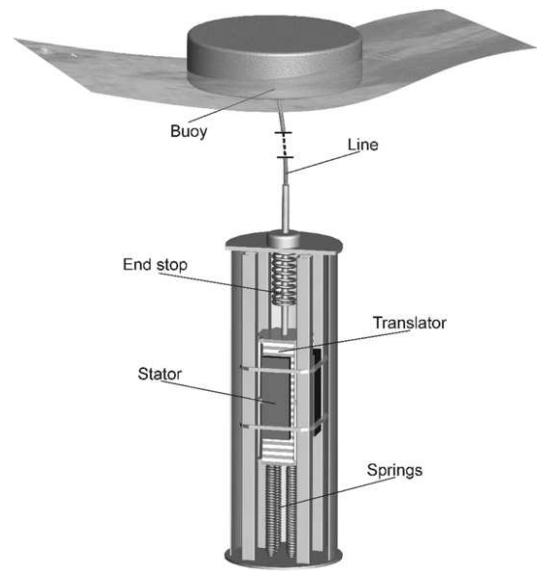}
	\caption{Schematic of the point-absorber type wave energy converter with a linear permanent magnet generator \cite{wang_nonlinear_2015}.}
	\label{fig:wec}
\end{figure}

As illustrated in Fig. \ref{fig:wec}, a PAWEC with a LPMG is used in this study. This device consists of a floating cylinder buoy that plays the role of capturing energy from ocean waves, a LPMG that plays the role of converting captured energy into electric energy, and a variable electric load that comsumes electric power generated by the LPMG. The diameter of the floating buoy is small compared to the incident wave length, thus this WEC belongs to the category of point absorber. The floating buoy is connected to the translator of a LPMG via steel wire and is activated by incident ocean waves. The buoy motion activates the translator to move up and down, which converts mechanical power into electric power. Since 2002, this technology has been developed by the wave energy research group at the Swedish Centre for Renewable Electric Energy Conversion at Uppsala University, and full-scale prototypes have been deployed at a site outside the town of Lysekil on the west coast of Sweden. In stand-alone cases, the LPMG is connected to an electric load after rectification by power electronic converters. Early work on validating this PAWEC concept via numerical analysis and real sea experiments are detailed in \cite{waters_experimental_2007,bostrom_experimental_2010}.

\begin{figure}[!ht]
	\includegraphics[width=7.8cm]{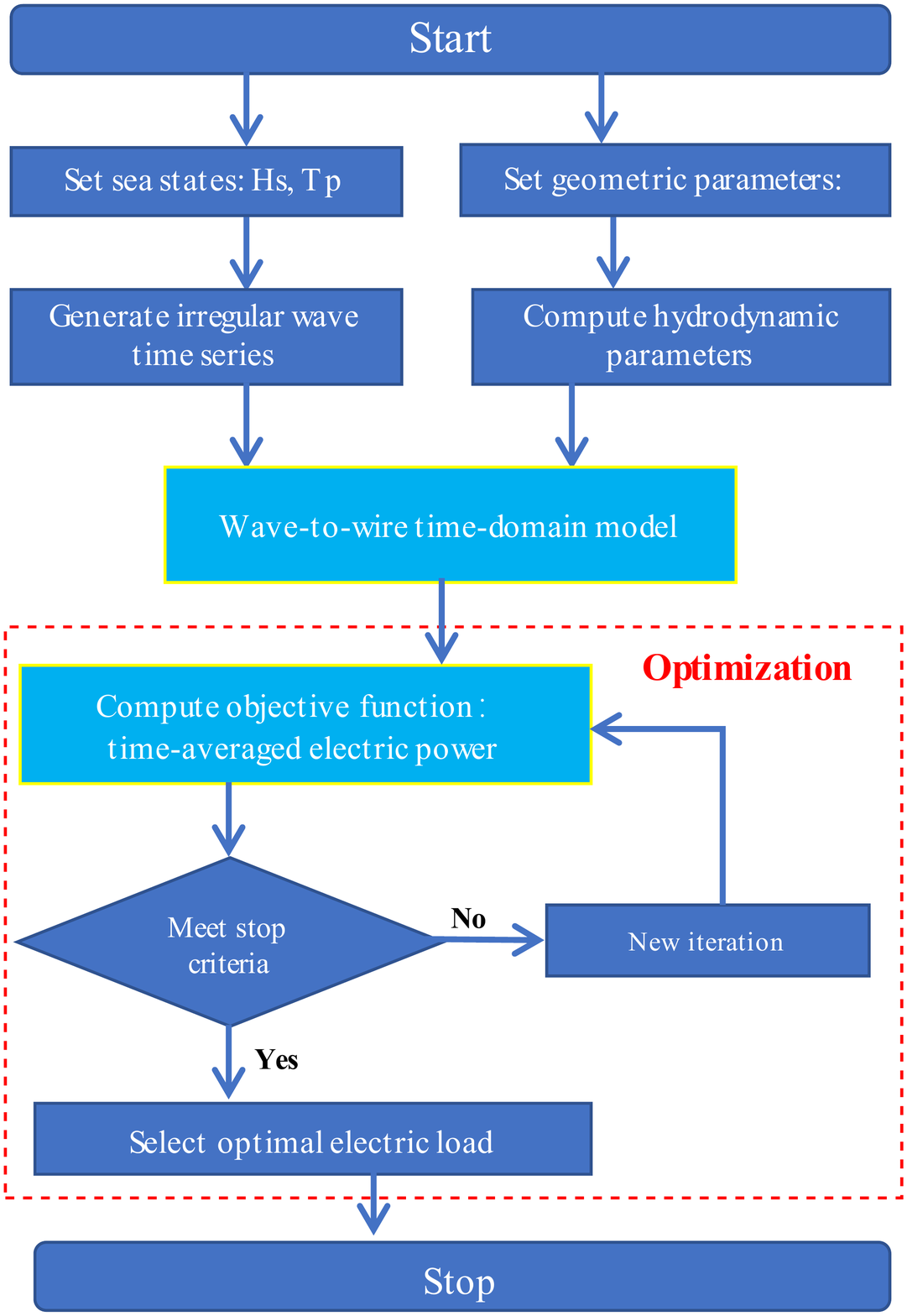}
	\caption{Procedure of optimizing electric load of a standalone wave energy converter using a wave-to-wire time-domain model.}
	\label{fig:methodology}
\end{figure}

This study aims to improve electric power generation, rather than power absorption, of an isolated PAWEC via electric load control. To identify the optimal electric load of a PAWEC in irregular waves, a wave-to-wire time-domain model and an optimization procedure are required. The problem of identifying the optimal electric load is formulated as a standard optimization problem, where the objective function is the time-averaged electric power that is computed by the developed wave-to-wire time-domain model, as  indicated in Fig. \ref{fig:methodology}.

\begin{figure}[!ht]
	\centering
	\includegraphics[width=7.8cm]{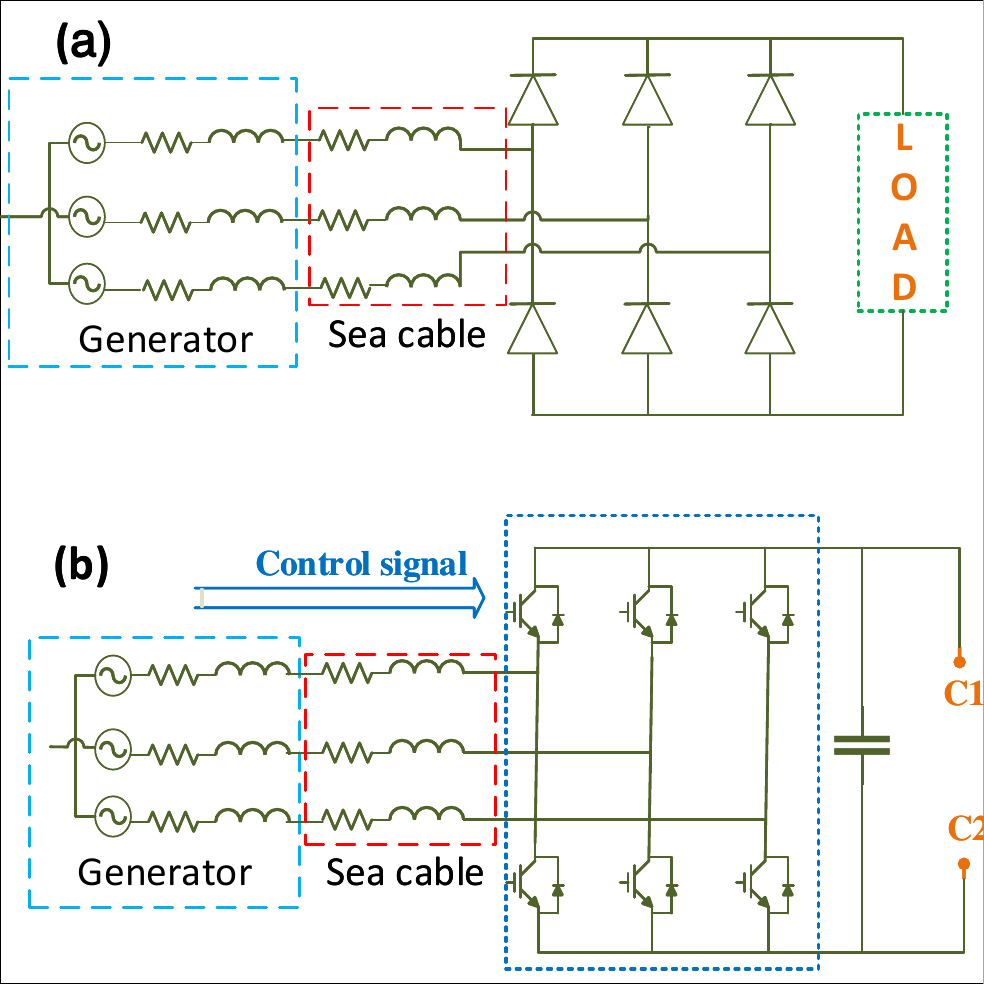}
	\caption{The three-phase circuit with passive rectification and synchronization rectification. (a) Passive rectification; (b) Synchronization rectification.}
	\label{fig:twoRectifications}
\end{figure}

Note that, wave-to-wire models depend, to a large extent, on the technologies used by a PAWEC, including the choice of converter topology, load, controllers and control schemes. Moreover, a specific controller can employ numerous control strategies. A mandatory associated process for a direct-driven LPMG is the power rectification, which is usually used to partially smoothen the variation of electric output of a LPMG due to the irregular nature of incident waves. Passive rectification and synchronous rectification are two widely used techniques. Passive rectification only supports unidirectional power flow and synchronization rectification allows for the bidirectional power flow. Fig. \ref{fig:twoRectifications}(a) shows a passive rectification using six diodes that converts the power from AC to DC, and Fig. \ref{fig:twoRectifications}(b) shows a synchronization rectification using insulated gate bipolar transistor (IGBTs).

\begin{figure}[!htbp]
	\centering
	\includegraphics[width=7.8cm]{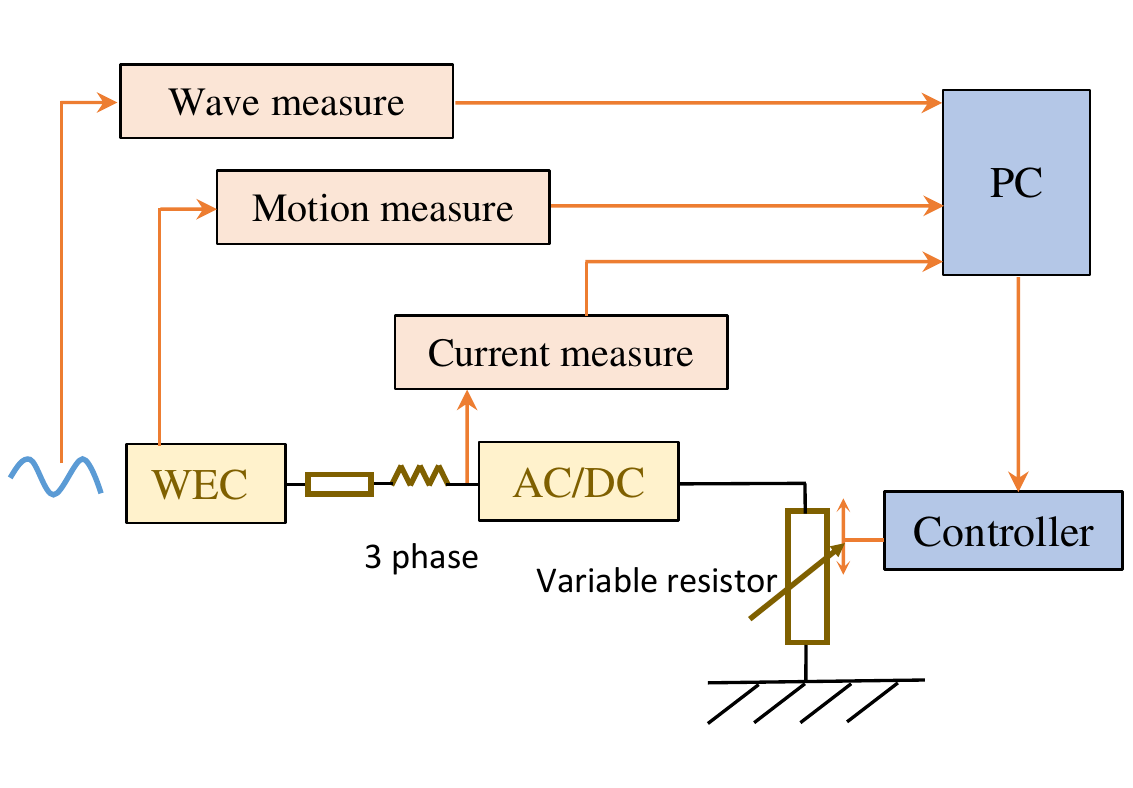}
	\caption{Control diagram used for the isolated point-absorber type wave energy converter with passive rectification. The variable resistor value is the parameter to be optimized and controlled.}
	\label{fig:controlDiagram}
\end{figure}

For an isolated PAWEC connected to a DC load, the passive rectification shown in Fig. \ref{fig:twoRectifications}(a) is employed, and the scheme used to control and optimize the electric power generated is to tune the value of the electric resistor load. Earlier experimental results have verified the influence of electric load on power generation \cite{waters_experimental_2007,bostrom_experimental_2010}. Fig. \ref{fig:controlDiagram} shows the control scheme for this passive rectification case, where the value of electric load can be tuned by the actuator of a controller, based on measured wave information and the PAWEC's motion. The tuning activity can be continuous or discontinuous, at a wave-to-wave level or sea-state-to-sea-state level, depending on the control need and user's requirements. It is assumed that the ideal actuator can achieve reference values quickly and the associated transient behavior is not discussed in this study.

The PAWEC has a diameter small compared to the wavelength, and is insensitive to wave direction, thereby only unidirectional incident waves are considered in the modeling and simulation. Additionally, as the LPMG works in heave direction, for simplicity, the equation of motion only considers the heave DOF.

\section{Wave-to-wire mathematical model of a WEC with a linear permanent magnet generator}
\label{sec:model}

\subsection{Regular and irregular incident waves}
Ocean waves can be generated by different mechanisms, e.g. by wind, earthquakes, or planetary forces. The majority are wind waves, driven by wind blowing over a surface area of the ocean. This study focuses on linear plane progressive waves, in regular form or irregular form.

For a regular plane-parallel wave, the variation of a surface wave at a space point is mathematically expressed as a cosine function:

\begin{equation}
\eta(x,t)= A_{wv} \cos(kx-\omega t + \theta),
\end{equation}
where $A_{wv}$ is wave amplitude, $\omega$ is angular frequency of incident waves, $t$ is time, $x$ is the space position in the direction of wave propagation, $k= 2 \pi / \lambda$ is the wave number with $\lambda$ representing the wavelength, and $\theta$ is the initial phase. 

However, most ocean waves are not regular, and the surface elevation varies irregularly over time and space. It is possible to describe the irregular waves by a combination of harmonic components. At one point in space, the time-continuous irregular wave elevation is defined as:
\begin{equation}
\xi(t) = \sum_{i=1}^N \alpha_i \cos(\omega_i t + \theta_i),
\end{equation}
where $\alpha_i$, $\omega_i$ and $\theta_i$ are the amplitude, angular frequency, and phase of the $i^{th}$ harmonic component, respectively.

The amplitude of harmonic wave $\alpha_i$ is obtained from the wave power spectrum, defined as:
\begin{equation}
\alpha_i = \sqrt{2S(\omega_i) d\omega}.
\end{equation}

To describe the ocean waves driven by wind in certain situations, several well-established spectrum can be used, e.g., the JONSWAP spectrum, the Pierson-Moskowitz spectrum, and the Bretschneider spectrum. For a fully developed sea, where a constant speed wind blows over a sufficiently long fetch of the ocean surface for a sufficiently long time, the Pierson-Moskowitz is commonly used. For a limited fetch, the JONSWAP spectrum is suggested. When a specific appropriate form of wave spectrum is well defined for average sea conditions, the Bretschneider spectrum can be used.

In this study, assuming the sea is fully developed, the Pierson-Moskowitz is employed. The Pierson-Moskowitz spectrum is defined by a peak wave period $T_p$ and a significant wave height $H_s$, formulated as:
\begin{equation}
S(\omega_i) = \frac{5\pi^4H_s^2}{T_p^4 \omega_i^5} \exp(-\frac{20\pi^4}{T_p^4 \omega_i^4}).
\label{eq:pmspectrum}
\end{equation}

\subsection{Wave-structure interaction}
Assuming that the fluid is homogeneous, incompressible, irrotational and inviscid, linear potential flow theory can be used to describe the interaction of a floating buoy with plane progressive waves in water of finite depth, with the free-surface and body-boundary conditions linearised. The assumption of linear potential flow permits the definition of a velocity potential $\Phi = Re(\phi e^{i\omega_{wv}t})$, yielding the Laplace equation in the fluid domain, i.e., $\Delta \Phi = 0 $. The linearised fluid-structure problem allows for the decomposition of velocity potential into incident, scatter and radiation components, i.e.:
\begin{equation}
{\phi} = \phi_I + \phi_S +  \phi_R.
\end{equation}

The complex velocity potential of the incident wave, $\phi_I$, is defined as:
\begin{equation}
\phi_I = \frac{igA_{wv}}{\omega_{wv}} \frac{\cosh[k(z+H_{dep})]}{\cosh{kH_{dep}}} e^{-ikx\cos \beta},
\end{equation}
where $\beta$ is the angle of incident wave relative to the positive direction of $x$ axis, $A_{wv}$ is the amplitude of the incident wave, $\omega_{wv}$ is the frequency of the incident wave, $H_{dep}$ is the water depth, $g$ is the gravitational acceleration, and $k$ is the real positive root of the dispersion relation $\omega_{wv}^2 = gk \tanh({kH_{dep}})$.

The excitation force due to incident waves is computed by integrating the hydrodynamic pressure over the wet surface $S_w$:

\begin{equation}
\hat{F}^i_e = - i \omega_{wv} \rho \int\int_{S_w} n_i (\phi_I + \phi_S) \mathrm{d}s,
\end{equation}
where the unit vector $\vec{\boldmath{n}} = (n_1, n_2, n_3)$ is normal to the body boundary and points out of the fluid domain.

For irregular waves, the excitation force in the heave direction can be represented as a summation of individual force component as:
\begin{equation}
F_e(t) = \sum_{i=1}^N \sigma (\omega_i) a_i(\omega_i) \cos(\omega_i t + \varphi_i),
\end{equation}
where $\sigma_i$ and $\varphi_i$ are the coefficient and phase of the excitation force under the $i^{th}$ harmonic wave, and can be calculated using, e.g., the commercial boundary element method solver WAMIT \cite{wamitWeb}.

The oscillation of a buoy in ocean water results in a radiation force, defined as:
\begin{equation}
F_r(t) = - m_{\infty} \ddot{z}(t) - \int_{-\infty}^t K(t-\tau) \dot{z}(t) d\tau,
\end{equation}
where $m_{\infty}$ is the infinite-frequency limit of added mass, and $K$ is the impulse response function, representing a memory of the oscillator velocity, and $K$ has a relationship with the added mass $A_{add}$ and the radiation resistance $B_{add}$ as follows:

\begin{equation}
A_{add}(\omega) - m_{\infty} = \frac{1}{\omega} \int_0^{\infty} K \sin(\omega t)d\omega,
\end{equation}

\begin{equation}
K(t) = \frac{2}{\pi} \int_0^{\infty} B_{add}(\omega) \cos(\omega t) d\omega.
\end{equation}

\subsection{Equation of motion of the primary capture system}
The PAWEC being developed at Uppsala University uses a linear permanent magnet synchronous generator. The natural wave motion is transferred to the generator's translator by rigid coupling. The translator moves up and down to generate electric power. For simplification, only the heave motion is considered in the mathematical modelling and simulation. For small wave heights, linear wave theory is used for the wave-absorber interaction, and it is assumed that the coupling between the absorber and translator is rigid without rope slacking. The motion of the WEC is described by Newton’s law as:

\begin{equation}
M_s \ddot{z}(t) = F_e(t) + F_r(t) + F_h(t) + F_{pto}(t),
\end{equation}
where $M_s$ is the inertial mass of the absorber and the generator translator, $z(t)$, $\dot{z}(t)$ and $\ddot{z}(t)$ denote vertical displacement, velocity, and acceleration of the absorber from equilibrium, respectively. $F_{pto}(t)$ is the PTO force, $F_h(t)$ is the hydrostatic restoring force due to the buoyancy and gravity, which is proportional to the displacement:
\begin{equation}
F_h(t) = -\frac{\rho g \pi D^2}{4} z(t),
\end{equation}
where $D$ is absorber's diameter.

\subsection{Linear permanent magnet generator}

\begin{figure}[!ht]
	\centering
	\includegraphics[width=7.8cm]{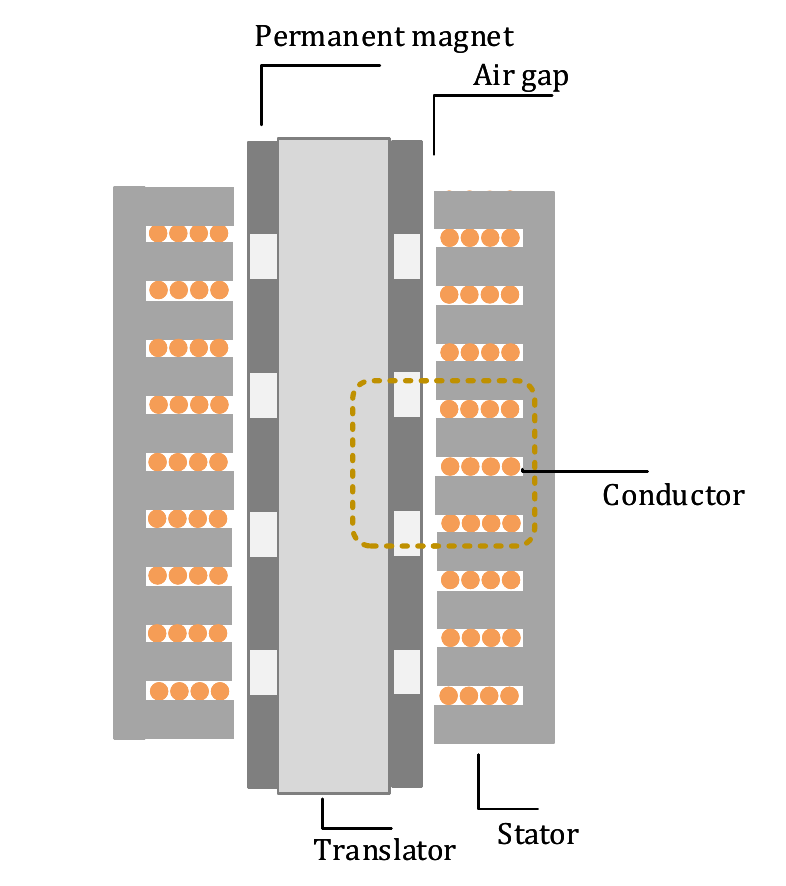}
	\caption{Illustration of the linear permanent magnet generator used in the wave energy converter.}
  \label{fig:generator}
\end{figure}

The power take-off system converts captured wave power into electric power. A LPMG is used as the PTO system of the PAWEC being developed at Uppsala University. As shown in Fig. \ref{fig:generator}, translator's surface is covered by permanent magnets that use Nd-Fe-B magnet materials with alternating polarity and separated by aluminium spacers. The stator consists of steel sheets and copper conductors that are wound in the stators slots. Magnetic flux variations are created when the translator moves relative to the stator, causing the induction of electric voltage. An electromagnetic force will be generated if the generator is connected to a closed electric circuit, and it counteracts the mechanical force on the translator.

The magnetic and electric phenomena in a LPMG can be described by field theory. The fundamental laws of electromagnetism are Maxwell's equation. The voltage induced in a closed circuit is proportional to the time rate of change of magnetic flux it encloses, which is a vector form of the Faraday's law of induction, i.e.:
\begin{eqnarray}
E = - N_c \frac{d \psi}{dt},
\end{eqnarray}
where $E$ is induced no-load voltage, also called electromotive force (EMF), $\psi$ is permanent magnetic flux, and $N_c$ is coil turns. For the 1-DOF PAWEC, the translator of a LPMG moves vertically, and the electromotive force can be written as follows,
\begin{equation}
E= - N_c \frac{d \psi}{dz} \frac{d z}{d t} = - N_c \frac{d \psi}{d z} \dot{z}(t).
\end{equation}

\subsection{Wave-to-wire integration}
To mathematically couple components' model in the integrated wave-to-wire simulation model, three assumptions are made for simplicity. 

The permanent magnetic flux $\psi$ is assumed to vary over the translator position, and is formulated as follows \cite{generatormodel2005}:
\begin{eqnarray}
\psi= \psi_{pm} \sin(\frac{2\pi}{\lambda_{pm}} x_{pm}(t) + \theta_{pm}),
\end{eqnarray}
where $x_{pm}$ is translator's position and varies over time, $\theta_{pm}$ is its initial phase, and $\lambda_{pm}$ is magnetic wavelength that is a function of pole width.

The variable $\psi_{pm}$ is the magnitude of magnetic flux, and it is determined by specific characteristics of a detailed LPMG. When the translator is fully exposed to the stator, in the case where translator's length is smaller than stator's length, the magnetic flux magnitude reaches a maximum value of $\psi_{pm,max}$; when the translator is completely out of the stator, the magnetic flux reaches it minimum values of zero; for other cases, a ratio factor $A_{act}$, corresponding to the active length of the translator, is defined to describe the ratio of $\lambda_{pm}$ and $\lambda_{pm,max}$, defined as follows:

\begin{equation}
A_{act}= \frac{1}{l_{st}} \lbrack \frac{1}{2}(l_{st} + l_{tr} - \vert x_{pm} \vert)\rbrack,
\end{equation}
where $l_{st}$ is length of the stator and $l_{tr}$ is length of the translator.

Another assumption is that, the total PTO force equals the electromagnetic force $F_{emf}$, written as
\begin{equation}
F_{emf} = \frac{3\pi}{2\tau_p} \lbrack i_d i_q (L_d- L_q) -\lambda_{pm} i_q \rbrack,
\end{equation}
where $i_q$, $i_d$, $L_q$ and $L_d$ are the quadrature-axes current, direct-axes current, quadrature-axes inductance and direct-axes inductance, respectively, and $\tau_p$ is the pole width of the permanent magnets. In the following simulations, considering that the quadrature-axes and direct-axes have the same inductance, only the quadrature-axes current $i_q$ contributes to the thrust.

The third assumption is that the coupling between the translator and the absorber is rigid in heave mode. In that way, velocity of the absorber exactly equals speed of the translator in heave mode.

\section{Numerical case Studies}
\label{sec:casestudy}
A full-scale PAWEC concept, being developed at Uppsala University, is used in the numerical case studies of Section \ref{subsec:case1} and Section \ref{subsec:optLoad}. The absorber has a diameter of $4 m$, a height of $1 m$ and a draft of $0.5 m$. The water depth is $20 m$. The pole width of permanent magnets is a value of $50 mm$, coil number is $10$ and magnetic flux magnitude $\psi_{pm}$ is $0.4 Wb$. The resistance and inductance of the generator are $0.5 \Omega$ and $11 mH$, respectively. The sea cable resistance is $0.5 \Omega$ and inductance is $1 mH$. The proposed wave-to-wire model is implemented in $MATLAB/Simulink$ environment, with a computation time step of $1 ms$. Hydrodynamic coefficients, e.g., added mass, radiation damping coefficients and excitation force coefficients, are computed using WAMIT, a boundary element code that uses the linear potential flow theory. Excitation force is estimated based on the predicted or pre-known incident wave information, while other approaches can be employed to approximate that, e.g. three approaches presented in \cite{Guo2018Numerical}.

\begin{figure}
	\centering
	\includegraphics[width=7.8 cm]{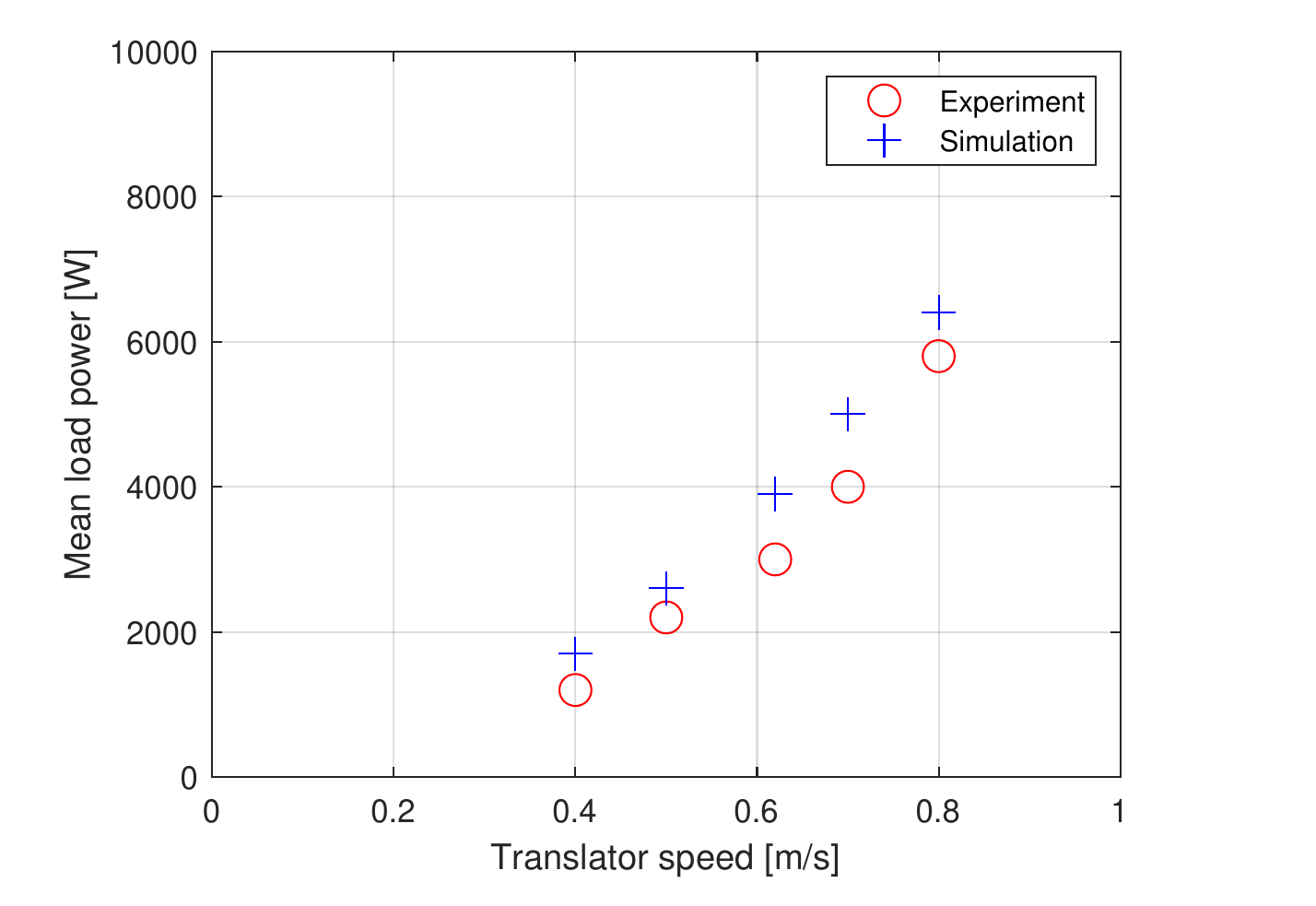}
	\caption{Simulated and experimental results of a linear permanent magnet generator that is connected to a $2.2 \Omega$ delta-connected resistive load.}
	\label{fig:generatorVerification}
\end{figure}

The simulation model of the LPMG is verified by results from physical experiments, as presented in Fig. \ref{fig:generatorVerification}. The generator is connected to a $2.2 \Omega$ delta-connected resistive load, and associated output current and voltage are measured and recorded to compute the electric power output. More experimental data and detailed measurements are presented in \cite{krishna_analysis_2013}, and information of experiment setup is described in \cite{rafael_thesis}.

\subsection{Case study: different load circuits}
\label{subsec:case1}
\begin{figure}[!h]
	\centering
	\includegraphics[width=7.8cm]{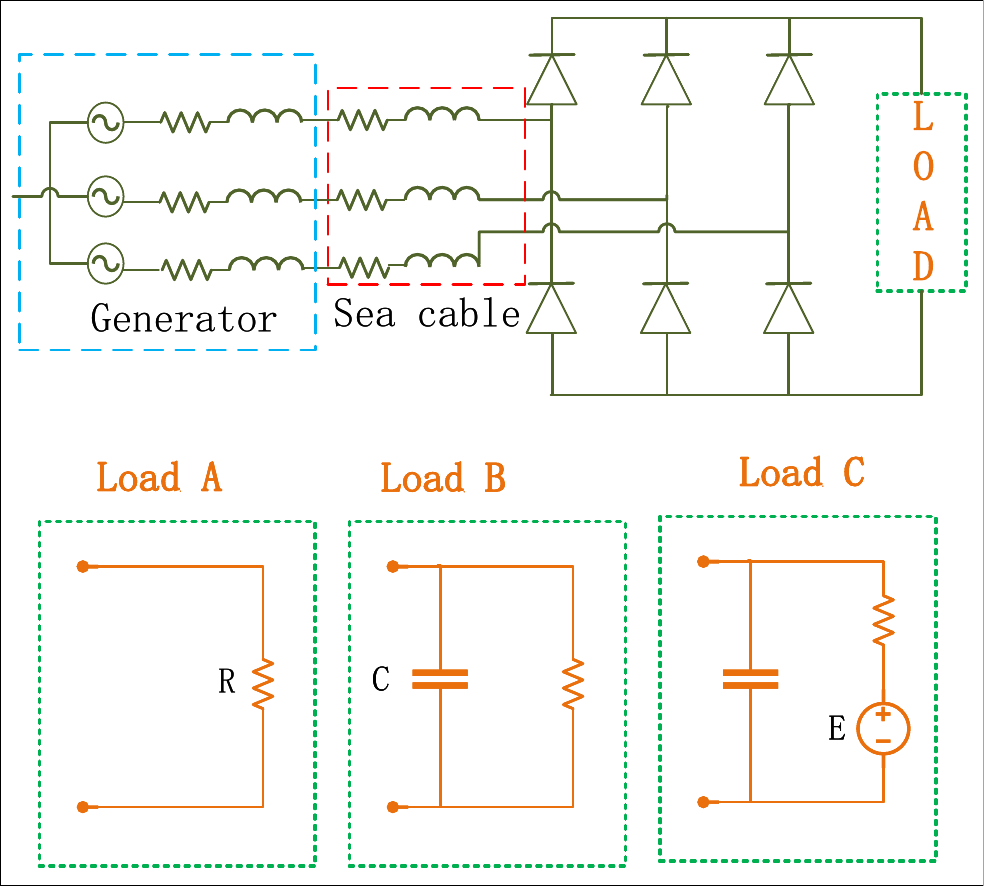}
	\caption{The three-phase circuit with passive rectification and different load types. Load A: a pure resistor. Load B: a resistor and a capacitor. Load C: a resistor, a capacitor and a DC voltage source. }
	\label{fig:loadTypes}
\end{figure}

This section examines the electric performance of a PAWEC under different DC loads, in regular and irregular waves. The LPMG is connected to a passive rectifier that uses six diodes to achieve the power conversion from AC to DC, as shown in Fig. \ref{fig:loadTypes}. Three different DC load circuits are connected to the passive rectifier, separately, i.e. Load A, Load B and Load C, as shown in Fig. \ref{fig:loadTypes}. These simple configurations serves as good reference cases for testing the influence of the electric load on the electric power generated.

Table \ref{table:powerof3loads} presents the time-averaged electric power, measured at the DC load side. Here the incident regular wave has an amplitude of $0.4 m$ and a frequency of $0.8 rad/s$. Results for Load A shows a strong influence of electric resistor load on the time-averaged electric power generated. Comparing the results from Load A and Load B indicates that the use of a parallel capacitor helps to increase power generation in the tested cases. Results for Load C indicate that the value of voltage source affects the electric power generated. Note that only a generator voltage higher than the DC voltage gives rise to a current through the passive rectifier.

\begin{table}[!htbp]
	\renewcommand{\arraystretch}{1.2}
	\label{table:powerof3loads}
\centering
	\small
	\begin{tabular}{p{0.20\columnwidth}p{0.06\columnwidth} p{0.10\columnwidth} p{0.08\columnwidth} p{0.18\columnwidth}}
		\hline
Load type & R $\lbrack \Omega \rbrack$ &  C \lbrack mF \rbrack &  E \lbrack V \rbrack &  Power \lbrack kW \rbrack \\
	\hline
	\hline
	Load A & 2  & N/A &N/A & 6.1 \\
	Load A & 10 & N/A & N/A & 3.8 \\
	Load A & 20 & N/A & N/A & 2.4 \\
	Load B & 2  & 60  & N/A & 6.2 \\	
	Load B & 10  & 60 & N/A & 3.9 \\	
	Load B & 20  & 60 & N/A & 2.6 \\
	Load C & 2   & 60 & 100 &  5.4\\
	Load C & 2  &  60 & 200 &  2.9\\
	Load C & 10  & 60 & 100 &  2.7\\
	Load C & 10  & 60 & 200 &  1.5\\
    \hline
	\end{tabular}
\caption{Time-averaged electric power generated by the LPMG when different load types and values are used.The incident regular wave has a height of $0.8 m$ and a frequency of $0.8 rad/s$.}
\end{table}

\begin{figure}[!ht]
	\centering
	\includegraphics[width=7.8cm]{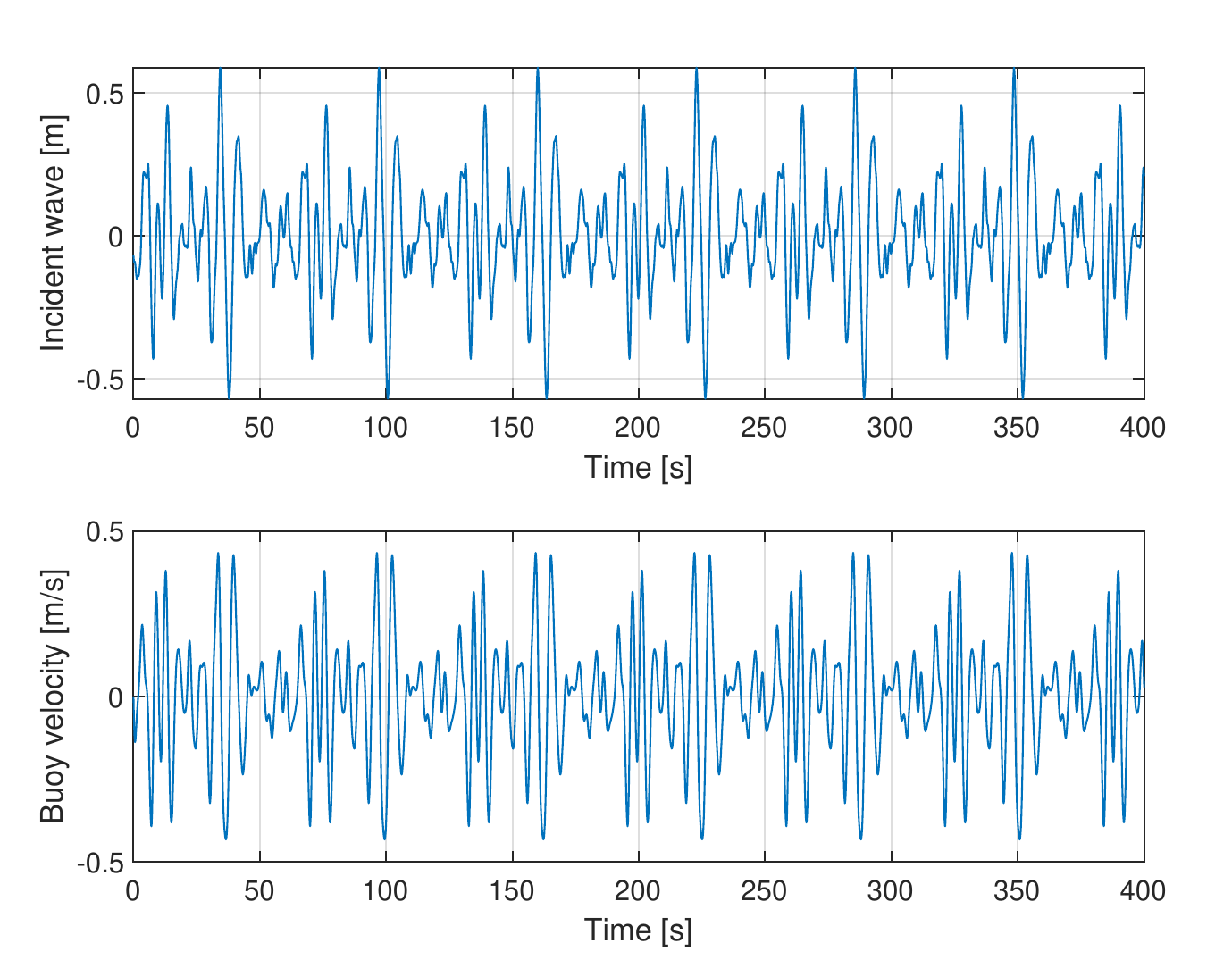}
	\caption{Velocity of the PAWEC under irregular waves. (a) Incident wave elevation; (b) Velocity. }
	\label{fig:motionIrr}
\end{figure}

\begin{figure}
	\centering
	\includegraphics[width=7.8cm]{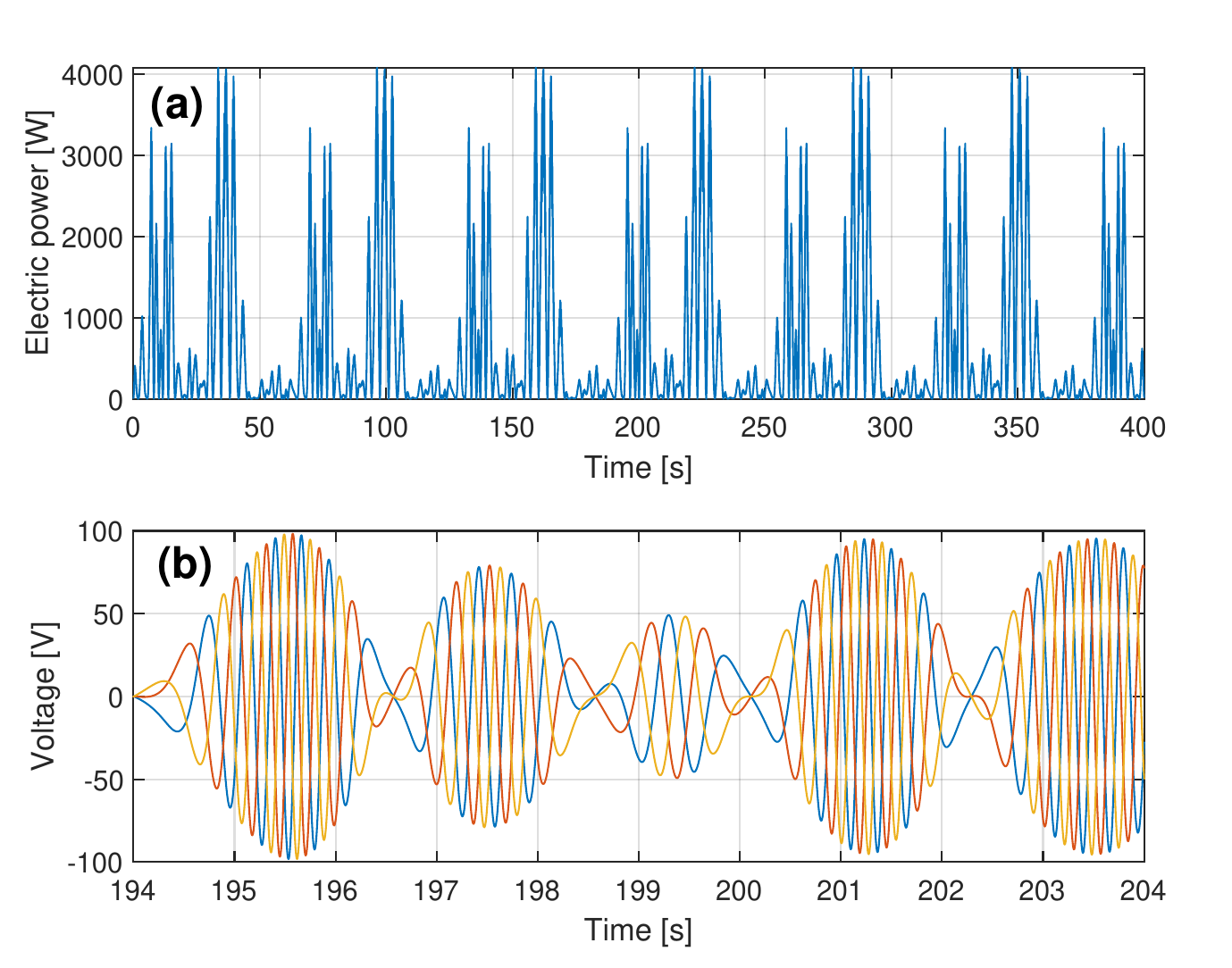}
	\caption{Output electric power and voltage of the PAWEC under irregular waves. (a) Instantaneous electric load power; (b) Three-phase no-load voltage during ten seconds.}
	\label{fig:eletricalPowerIrr}
\end{figure}

Additionally, dynamics of the PAWEC in irregular waves is investigated using the developed wave-to-wire model. Fig. \ref{fig:motionIrr} presents the resultant velocity of the PAWEC in an irregular wave profile, created by the Pierson-Moskowitz spectrum defined in eq.(\ref{eq:pmspectrum}) with a peak wave frequency of $0.8 rad/s$ and a significant wave height of $0.8 m$. The passive rectifier is connected to a resistor that has a value of $3 \Omega$. In Fig. \ref{fig:eletricalPowerIrr} the instantaneous load power and no-load voltage are plotted. This shows that the instantaneous electric power consumed by the resistor load varies in amplitude, and its peak amplitude is six times higher than the time-averaged value. Fig. \ref{fig:eletricalPowerIrr}(b) shows the no-load voltage in a $10 s$ time window, and this three-phase voltage varies in amplitude and frequency.

\subsection{Case study: Optimal load estimation}
\label{subsec:optLoad}
This section aims to find the optimal electric load for the PAWEC under specific wave conditions. This is an maximization problem where the variable to be optimized is the electric load, i.e. resistor value is this case, and the objective function is the time-averaged electric power consumed by the electric load, defined as:
\begin{equation}
\overline{P_{dc}} = \int_{0}^{T_{mn}} V_{dc}I_{dc} dt,
\end{equation}
where $V_{dc}$ and $V_{dc}$ are the voltage and current measured at the electric load, respectively, and $T_{mn}$ is the total time used in the simulation. 

\begin{figure}[!ht]
	\centering
	\includegraphics[width=7.8 cm]{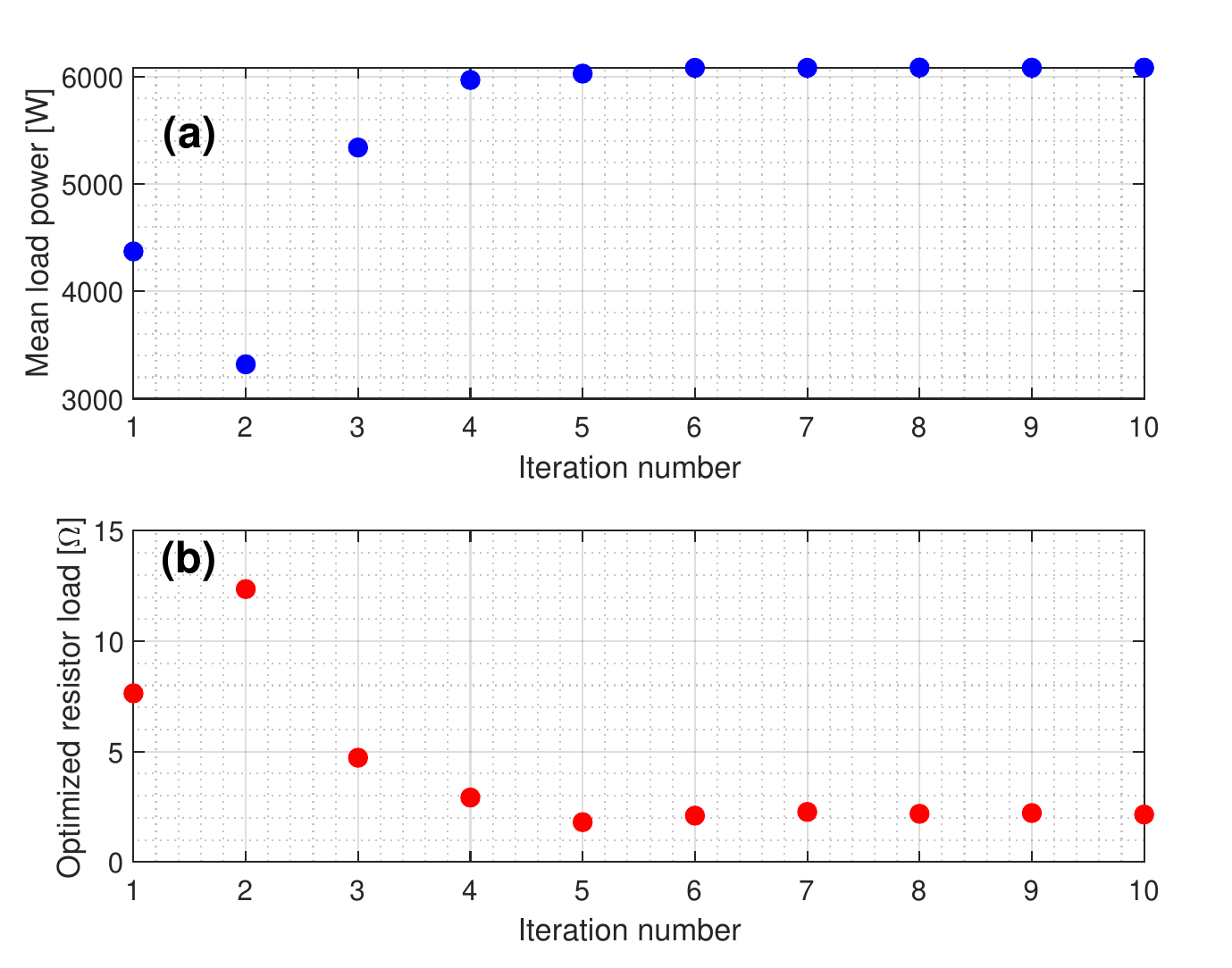}
	\caption{Evolution of the optimization process. (a) Time-averaged load power; (b) Optimized resistor load value. Wave height is $0.8 m$ and wave frequency is $0.8 rad/s$.}
	\label{fig:optimization}
\end{figure}

The optimization procedure is depicted in Fig. \ref{fig:methodology}. A golden-section search algorithm \cite{KHELDOUN2016125} is used to solve the optimization problem, and the developed wave-to-wire model is implemented in $MATLB/Simulink$ that simulates and computes the time-averaged power in time domain (with a time step of $1 ms$). The passive rectifier is connected to a resistor, i.e. Load A shown in Fig. \ref{fig:loadTypes}. Fig. \ref{fig:optimization} presents the optimized resistor value and the resulting load power at each optimization iteration. Results indicate that the optimal electric load value can be found efficiently and the time-averaged load power is significantly improved compared to the initial case (i.e., the first iteration). 

\section{Discussion}
\label{sec:discussion}
Sections \ref{sec:casestudy} demonstrates the application of a time-domain wave-to-wire model to a PAWEC with a LPMG. Through numerical investigations, the influence of the electric load on electric power generation is demonstrated. Additionally, the developed optimization procedure that employs the proposed wave-to-wire model can efficiently search for the optimal electric load for a PAWEC under specific wave conditions. This has strong implications for design of load controllers, as variable speed generators are used more and more frequently during the conversion from ocean wave source into electric power, where generator output voltage varies both in amplitude and frequency and the electric load connected to a LPMG should be actively tuned according to the change of sea states. If the computation time of the optimization procedure is faster than the change of sea states, the proposed optimal-load searching method can be used for load control of full-scale WECs in physical experiments and during operation.

The objective of this study is to increase the electric power measured at the load side, rather than the absorber power that is used as the objective function in most control strategy studies. Indeed, the difference between them is huge, for example, the time-averaged mechanical power of the LPMG is $11 kW$ in the specific case of Section \ref{subsec:optLoad} while the time-averaged load power is $6.1 kW$. This is due to the losses of electric power cable, as well as characteristics of a specific LPMG. The influence of non-ideal PTO on the energy conversion was also investigated in other studies, e.g., in \cite{genest_effect_2014}. Thereby, in an electric wave power system, it is necessary to account for the increased losses and not only look at the increase in absorbed power.

Additionally, the wave-to-wire model based optimization procedure that only optimize the resistor load value in the case study can be applied to other load types with more variables to be optimized. In a case with more variables, the golden-section search algorithm can be easily replaced by other multiple-variable optimization algorithms, e.g., a genetic algorithm.

\section{Conclusions}
	\label{sec:conclusion}
This paper presents a generic methodology for optimizing electric load of a wave energy converter with a linear permanent magnet generator. The demonstrated wave-to-wire mathematical model helps in understanding fundamental electric dynamics of the one-degree-of-freedom device, and the developed optimization routine can efficiently find the optimal resistor load value for a device under regular waves or irregular waves. The optimization procedure is applied to a full-scale point-absorber type wave energy converter that is being developed at Uppsala University. Numerical results indicate that the performance of the device can be significantly improved, in terms of energy generation. 

The output voltage from a point-absorber type wave energy converter varies in frequency and amplitude. For an off-grid case, passive rectification is required for a DC load. It is concluded that the electric load of a standalone wave energy converter should be actively tuned according to the change of sea states, which can significantly increase the electric power generated. The passive-control based optimization routine can be applied to real-time operation of a full-scale wave energy in random waves.

\section{Acknowledgments}	
This work was supported by the Sun Yat-Sen University (grant number 76170-18841210); Swedish Research Council (grant number 40421-1); and Liljewalchs travel scholarships.

\bibliographystyle{elsarticle-num} 
\bibliography{new.bib}
\end{document}